\documentclass[aps, physrev, twocolumn, superscriptaddress,longbibliography]{revtex4-2}

\usepackage{graphicx}%
\usepackage[colorlinks=true,linkcolor=blue,urlcolor=blue,citecolor=blue]{hyperref}
\usepackage{nameref}
\usepackage{siunitx}
\usepackage{amsmath}
\usepackage{xcolor}
\usepackage{natbib}

\begin{document}


\title{Angle-Dependent Magnetoresistance Induced by Interface-Generated Spin Current in RuO$_2$/Permalloy Heterostructures}


\author{Akashdeep Akashdeep}
\email[]{aakashde@uni-mainz.de}
\affiliation{Institute of Physics, Johannes Gutenberg-University Mainz, 55128 Mainz, Germany}

\author{Ewiese Mohammad Ababneh}
\affiliation{Department of Physics, Indiana University, Indianapolis, Indiana 46202, USA}

\author{Christin Schmitt}
\affiliation{Institute of Physics, Johannes Gutenberg-University Mainz, 55128 Mainz, Germany}

\author{Edgar Galíndez-Ruales}
\affiliation{Institute of Physics, Johannes Gutenberg-University Mainz, 55128 Mainz, Germany}

\author{Felix Fuhrmann}
\affiliation{Institute of Physics, Johannes Gutenberg-University Mainz, 55128 Mainz, Germany}

\author{Timo Kuschel}
\affiliation{Institute of Physics, Johannes Gutenberg-University Mainz, 55128 Mainz, Germany}
\affiliation{Center for Spinelectronic Materials and Devices, Department of Physics, Bielefeld University, 33615 Bielefeld, Germany}

\author{Mathias Kläui}
\email[]{klaeui@uni-mainz.de}
\affiliation{Institute of Physics, Johannes Gutenberg-University Mainz, 55128 Mainz, Germany}
\affiliation{Department of Physics, Center for Quantum Spintronics, Norwegian University of Science and Technology, 7491 Trondheim, Norway}

\author{Vivek Amin}
\email[]{vpamin@iu.edu}
\affiliation{Department of Physics, Indiana University, Indianapolis, Indiana 46202, USA}

\author{Gerhard Jakob}
\email[]{jakob@uni-mainz.de}
\affiliation{Institute of Physics, Johannes Gutenberg-University Mainz, 55128 Mainz, Germany}



\begin{abstract}
Altermagnets, a recently discovered class of magnetic materials exhibiting ferromagnetic-like spin-split bands and antiferromagnetic-like compensated magnetic order, have attracted significant interest for next-generation spintronic applications. Ruthenium dioxide (RuO\textsubscript{2}) has emerged as a promising altermagnetic candidate due to its compensated antiparallel magnetic order and strong spin-split electronic bands. However, recent experimental and theoretical reports also suggest that RuO\textsubscript{2} may be non-magnetic in its ground state, underscoring the need for deeper investigations into its magnetic character. Specifically, the (100)-oriented RuO\textsubscript{2} films are expected to generate spin currents with transverse spin polarization parallel to the Néel vector. Here, we investigate magnetotransport in epitaxial RuO\textsubscript{2}/Permalloy (Py) heterostructures to examine spin Hall magnetoresistance and interfacial effects generated in such systems. Our measurements reveal a pronounced negative angular-dependent magnetoresistance for variation of magnetic field direction perpendicular to the charge current direction. Detailed temperature-, magnetic field-, and crystallographic orientation-dependent measurements indicate that interface-generated spin current (IGSC) at the RuO\textsubscript{2}/Py interface predominantly governs the observed magnetoresistance. In addition, the role of IGSC contribution to the observed magnetoresistance was demonstrated through drift-diffusion calculations. This shows that strong interface effects dominate over possible altermagnetic contributions from RuO\textsubscript{2}. Our results show that the role of interface-generated spin currents is crucial and should not be overlooked in studies of altermagnetic systems. A critical step in this direction is disentangling interfacial from altermagnetic contributions. The insight into interfacial contributions from altermagnetic influences is essential for the advancement of RuO2-based spintronic memory and sensing applications.

\end{abstract}


\maketitle

\section{Introduction}
The field of magnetism has long been a cornerstone of condensed matter physics, with ferromagnetism and antiferromagnetism as the two traditionally recognized magnetic phases characterized by collinear spin arrangements. Recently, a new class of magnetic materials known as altermagnets has emerged, exhibiting unique properties that challenge conventional understanding. Altermagnets are characterized by a collinear, compensated magnetic order similar to antiferromagnets (AFMs) but also feature spin-split energy bands, a hallmark of ferromagnets (FMs). This unique combination makes them highly promising for spintronic applications, as they enable efficient spin current generation without net magnetization and reduction of stray fields \cite{Smejkal2022_NatRevMat, Smejkal2022_PhysRevX, Bose2022}, overcoming some of the limitations associated with traditional FMs and heavy metals (HMs) in spin-based memory and logic devices. Spin current can be generated through multiple mechanisms, including the spin Hall effect (SHE) \cite{PhysRevLett.83.1834}, the Rashba-Edelstein effect (REE) \cite{EDELSTEIN1990233},  and interface-generated spin current (IGSC) \cite{PhysRevLett.121.136805}. Among these, IGSC plays a pivotal role in FM/non-magnetic heterostructures, which govern spin-charge conversion and interfacial transport effects that are critical for spintronic applications.  Understanding how IGSC interacts with novel materials such as altermagnets is essential for developing next-generation spintronic technologies.

One material that has emerged as a promising candidate in the field of altermagnetism is ruthenium dioxide (RuO\textsubscript{2}). RuO\textsubscript{2}, with its rutile crystal structure, has been shown to possess a compensated antiparallel sublattice magnetic order \cite{Berlijn2017_PhysRevLett, Zhu2019_PhysRevLett}, and also exhibits strong spin splitting in its electronic band structure \cite{Feng2022_NatElectron, Bai2022_PhysRevLett, Karube2022_PhysRevLett}. This material has drawn significant attention in spintronics because of its potential to generate an anomalous Hall effect (AHE) signal and spin currents \cite{Feng2022_NatElectron, Bose2022}. The Néel vector in RuO\textsubscript{2}, which defines the direction of the magnetic order, can be reoriented from the c-axis by applying a strong magnetic field \cite{Smejkal2022_NatRevMat}. Its altermagnetic phase is further associated with a unique spin-momentum locking mechanism \cite{Smejkal2022_PhysRevX, PhysRevLett.126.127701}, making it a valuable platform for spin-current manipulation. Importantly, it has been shown that the RuO\textsubscript{2} spin-splitting and spin current generation strongly depend on the crystallographic orientation \cite{Bai2022_PhysRevLett}. For instance, a (100)-oriented RuO\textsubscript{2} film can generate a spin current with transverse spin polarization parallel to the Néel vector [001] when a charge current is applied along the axis [010] \cite{Smejkal2022_NatRevMat}. These properties make RuO\textsubscript{2} a strong candidate for energy-efficient spintronic devices, such as next-generation magnetoresistive memory and spin-orbit torque-based switching technologies. Investigating spin Hall magnetoresistance (SMR) in RuO\textsubscript{2}/FM heterostructures is particularly relevant for practical applications in magnetic sensing and memory technologies, for which interfacial spin transport plays a crucial role \cite{PhysRevLett.110.206601, PhysRevLett.121.136805}. In these systems, interface spin-orbit coupling fundamentally governs spin current reflection and absorption, mediating magnetoresistance effects. Recent experimental and theoretical reports have also indicated that RuO\textsubscript{2} might be non-magnetic in its ground state \cite{PhysRevLett.132.166702, kessler2024absence, PhysRevB.109.134424}. This apparent contradiction underscores the need for further detailed investigation to clarify the magnetism of RuO\textsubscript{2} and to deepen our understanding of its predicted rich spintronic properties.

Given the unique spin-polarized transport properties of RuO\textsubscript{2} \cite{PhysRevLett.126.127701}, and our previous samples showing an altermagnetic signature \cite{doi:10.1126/sciadv.adj4883, weber2024all}, this study investigates RuO\textsubscript{2}/FM heterostructures to explore how altermagnetism and IGSC contribute to SMR. While SMR is typically studied using HMs as the non-magnetic layer, we investigate whether RuO\textsubscript{2} can act as a spin source rather than an HM exhibiting SMR in FM heterostructures. Here, we explore the magnetotransport properties of RuO\textsubscript{2}/permalloy (Py) heterostructures. Our magnetotransport measurements probe the interface between altermagnetic RuO\textsubscript{2} and FM Py to investigate if the interplay of the altermagnetic nature of RuO\textsubscript{2} and the FM properties of Py can be used to generate a magnetoresistance based on the spin-polarized current in RuO\textsubscript{2}. We further explored the role of IGSC in conjunction with inverse SHE (iSHE) and demonstrated its contribution to the observed magnetoresistance through drift-diffusion calculations. Understanding spin-polarized transport at RuO\textsubscript{2}/FM interfaces could reveal new pathways for controlling magnetoresistance, providing crucial insights for developing energy-efficient spintronic memory and logic devices.

\section{Methods}
Epitaxial rutile RuO\(_2\)(100) films with a thickness of 5 and \SI{8}{\nano\meter} were grown on TiO\(_2\)(100) substrates using pulsed laser deposition in an ultrahigh vacuum chamber maintained at a base pressure below \SI{2e-8}{\milli\bar}. A krypton fluoride excimer laser (wavelength \SI{248}{\nano\meter}) operating at a pulse energy of \SI{130}{\milli\joule} and a frequency of \SI{10}{\hertz} was employed for the ablation process. The deposition was conducted under a controlled oxygen atmosphere of \SI{0.02}{\milli\bar} at a substrate temperature of \SI{400}{\celsius}. Post-deposition, the films were cooled to room temperature at a controlled rate of \SI{25}{\kelvin\per\minute}. In-situ reflection high-energy electron diffraction (RHEED) was used to monitor the crystalline structure during growth. The presence of the RHEED pattern indicates crystallinity extending to the surface. The RuO\(_2\) samples were transferred to a sputter deposition chamber using a vacuum transport chamber to prevent contamination at the RuO\(_2\)/Py interface. A \SI{4}{\nano\meter} layer of Py and a \SI{2}{\nano\meter} Aluminum capping layer were deposited at room temperature under an Argon atmosphere maintained at \SI{0.02}{\milli\bar}. The Al capping naturally oxidized and did not contribute to current transport in later experiments. Six terminal Hall bar devices of \SI{20}{\micro\meter} width and \SI{100}{\micro\meter} length  were fabricated using optical lithography. The distance between voltage probes for measurement was \SI{50}{\micro\meter}. Two sets of Hall bar devices were fabricated for longitudinal resistance measurements and were aligned along the in-plane RuO\(_2\)[001] (c-axis) and RuO\(_2\)[010] crystallographic axes of the RuO\(_2\)(100) films to apply charge current (\( J_C \)) along the exact crystallographic directions to enable directionally resolved magnetotransport measurements sensitive to spin current polarization. Measurements were conducted in a delta mode configuration using a Keithley 6221 sourcemeter and Keithley 2182 nanovoltmeter. A 3D vector cryostat was employed to perform temperature-dependent measurements down to \SI{20}{\kelvin}. In the angular-dependent magnetoresistance (ADMR) scans, a fixed magnetic field of $\mu_0 H = \SI{0.95}{\tesla}$ was rotated in specific planes. Control samples of Py on TiO\(_2\) have been prepared under identical conditions but omit the RuO\(_2\) layer in order to separate the anisotropic magnetoresistance (AMR) effects of Py.

\section{Results}
\subsection{Structural and Surface Analysis}
\begin{figure}
    \centering
    \includegraphics[width= \columnwidth]{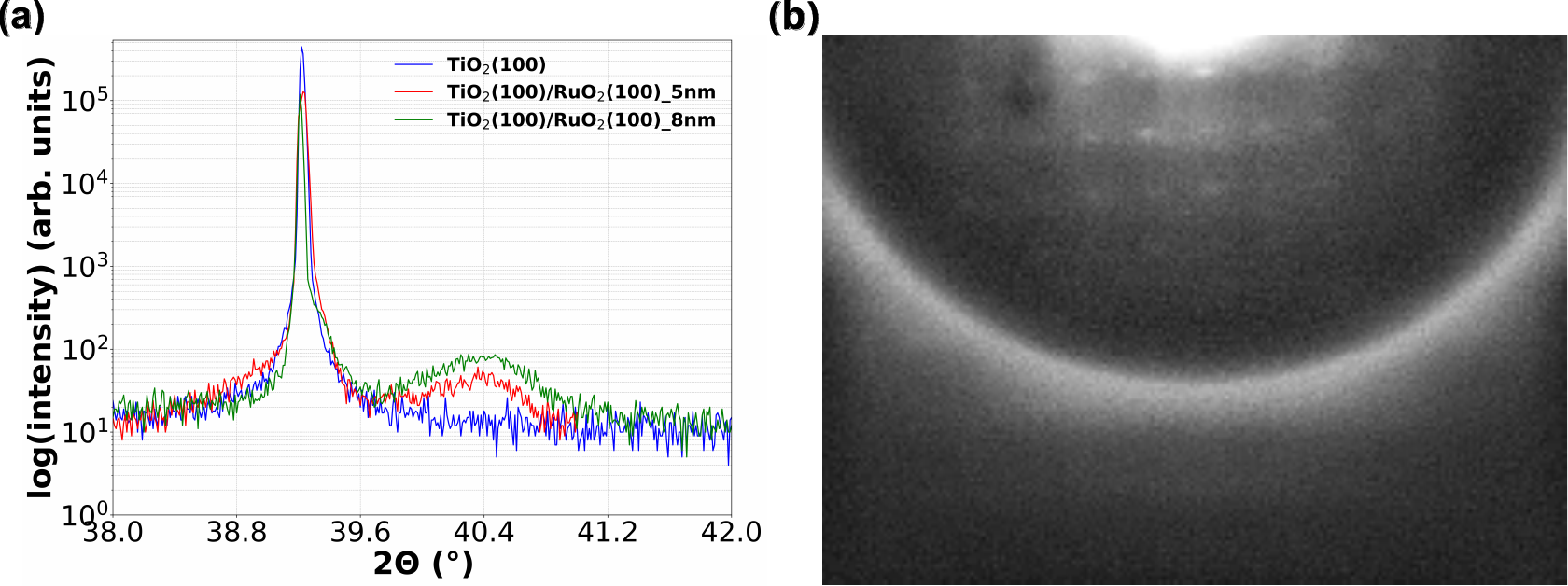} 
    \caption{Structural and surface characterization of TiO\(_2\)(100)//RuO\(_2\)(100)/Py films. (a) XRD patterns measured with the scattering vector normal to the (100)-oriented rutile substrate. (b)  RHEED pattern for TiO\(_2\)(100)//RuO\(_2\)(100), with the electron beam aligned along TiO\(_2\)[001].}
    \label{fig_1}
\end{figure}

\sloppy
Figure~\ref{fig_1} provides an analysis of the structural and surface characteristics of the TiO\textsubscript{2}(100)//RuO\textsubscript{2}(100)/Py films. Figure~\ref{fig_1}(a) shows the X-ray diffraction (XRD) patterns measured with the scattering vector normal to the (100)-oriented rutile substrate. The epitaxial growth of RuO\(_2\) films on the TiO\(_2\) rutile substrate is clearly evident. The diffraction feature of RuO\(_2\)(100) is weak due to the thin film's low thickness, but the diffraction features become stronger as the film's thickness increases compared to the bare substrate. Figure~\ref{fig_1}(b) presents the RHEED pattern for the TiO\(_2\)(100)//RuO\(_2\)(100) system, obtained with the electron beam aligned along TiO\(_2\)[001]. The transmission diffraction pattern corroborates the crystalline growth indicated by the XRD results but reveals that the surface is not atomically flat \cite{kokosza2021simplified}. The extended structural and transport characterization is provided in the supplemental material \cite{SM_IGSC_RuO2Py}.

\subsection{Comparison of Magnetoresistance in Py to RuO\(_2\)/Py}

\begin{figure*}
    \centering
    \includegraphics[width=\textwidth]{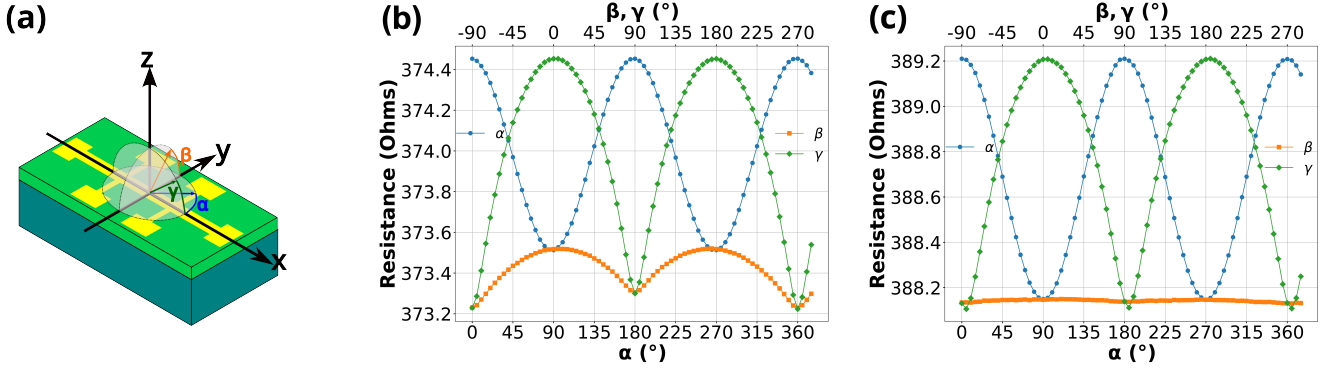} 
    \caption{Longitudinal resistance variations for magnetic field direction in the $\alpha$-, $\beta$-, and $\gamma$-planes. (a) Measurement scheme for the $\alpha$-, $\beta$-, and $\gamma$-planes (b) RuO$_2$(100) [\SI{5}{\nano\meter}]/Py heterostructure, and (c) Py thin film measured at \SI{20}{\kelvin}, for \( J_C \) being parallel to the c-axis under a \SI{0.95}{\tesla} magnetic field.}
    \label{fig_2}
\end{figure*}

The longitudinal resistance of the RuO\textsubscript{2}/Py heterostructure was measured while varying the Py magnetization direction in three distinct planes (\(\alpha\), \(\beta\), and \(\gamma\)) at temperatures of \SIlist{20;250;300}{\kelvin}. Figure~\ref{fig_2}(a) provides a schematic of the measurement scheme. The \(\alpha\)-plane (xy-plane) corresponds to the case for which the external magnetic field is rotated in-plane, crossing both parallel ($\alpha = \SI{0}{\degree}$) and perpendicular directions relative to the charge current. In this configuration, the relative resistance change of the RuO\textsubscript{2}[\SI{5}{\nano\meter}]/Py bilayer was $\sim \SI{-3e-3}{}$, calculated as \( {(R - R(0^\circ))}/{R(0^\circ)} \), for the \SI{20}{\kelvin} case. The \(\gamma\)-plane (xz-plane) involves rotating the external magnetic field from an in-plane orientation, for which $\gamma = \SI{0}{\degree}$ (parallel to the charge current), to an out-of-plane configuration, resulting in a resistance change of $\sim \SI{-4e-3}{}$ at \SI{20}{\kelvin}. In the \(\beta\)-plane (yz-plane), the field was rotated from an in-plane orientation, for which $\beta = \SI{0}{\degree}$ (normal to the charge current), to an out-of-plane position, resulting in a relative resistance change of $\sim \SI{-1e-3}{}$ at \SI{20}{\kelvin}. Figures~\ref{fig_2}(b) and \ref{fig_2}(c) illustrate these resistance variations for RuO\textsubscript{2}[\SI{5}{\nano\meter}]/Py and bare Py films, respectively, at \SI{20}{\kelvin}.

\begin{figure*}
    \centering
    \includegraphics[width= \textwidth]{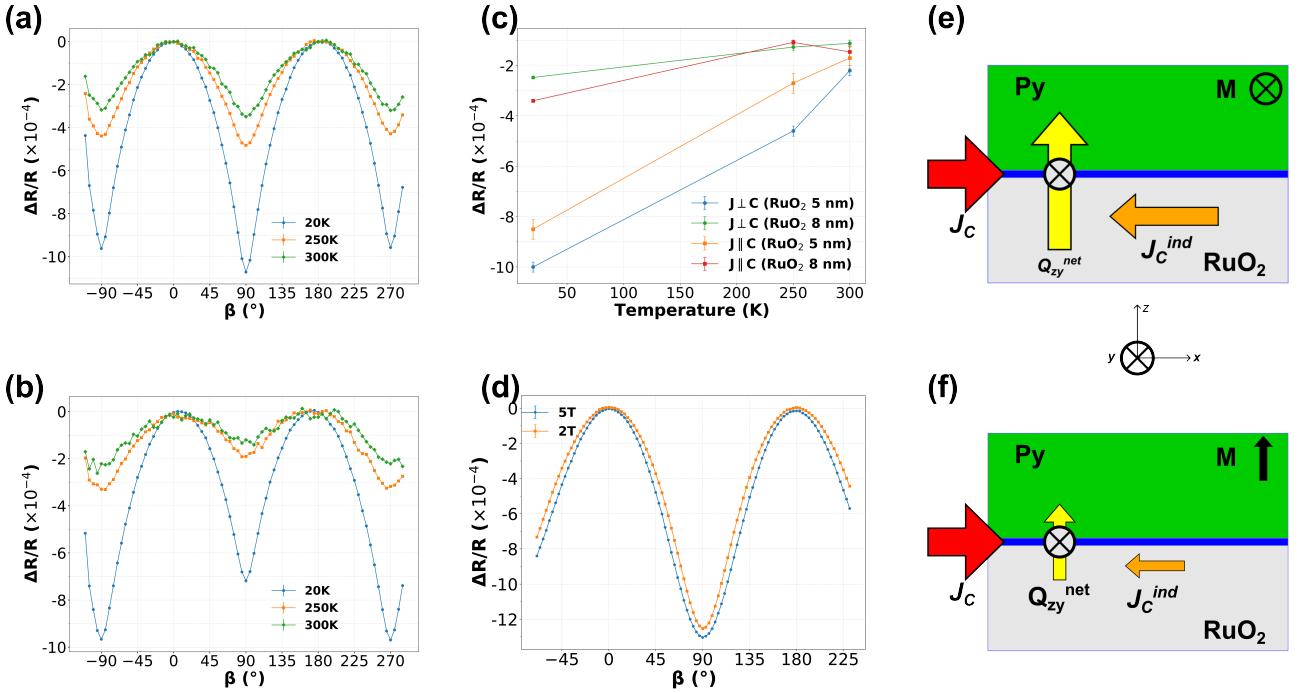} 
    \caption{Longitudinal resistance variations in the \(\beta\)-plane at different temperatures for (a) \( J_C \perp c \) and (b) \( J_C \parallel c \) configurations for RuO\(_2\)[\SI{5}{\nano\meter}]/Py under a \SI{0.95}{\tesla} magnetic field. (c) Temperature dependence of the peak-to-peak magnetoresistance amplitude in the \(\beta\)-plane. (d) ADMR measured at \SI{2}{\tesla} and \SI{5}{\tesla} for RuO\(_2\)[\SI{5}{\nano\meter}]/Py  for \(J_C \parallel c\) in the \(\beta\)-plane at \SI{20}{\kelvin}. (e, f) Schematic representation of IGSC-induced negative ADMR. When a charge current \( J_C \) (red arrow) is injected, a spin current \( Q_{zy} \) (yellow arrow) is generated perpendicular to the interface. The magnitude of the spin current \( Q_{zy} \) is maximized and minimized when the magnetization \( M \) (black arrow) is (e) parallel and (f) perpendicular to the spin index on the y-axis, respectively. Consequently, the inverse process of SHE converts the interfacial spin current back into an induced charge current \( J_C^{\text{ind}} \) (orange arrow), resulting in a resistance change.
}
    \label{fig_3}
\end{figure*}

At an elevated temperature of \SI{300}{\kelvin}, resistance changes for magnetic field direction in the \(\alpha\)- and \(\gamma\)-planes were reduced to half their amplitudes measured at \SI{20}{\kelvin}, while a substantial reduction (to one-fifth) was observed for magnetic field direction in the \(\beta\)-plane as shown in Fig.~\ref{fig_3}(c). The ADMR for magnetic field directions in the \(\beta\)-plane has a phase shift of \SI{90}{\degree} compared to SMR in the FM/HM system. In bare Py films grown on TiO\(_2\), ADMR in the \(\alpha\)- and \(\gamma\)-planes have same dependence as of RuO\(_2\)/Py. However, the \(\beta\)-plane response was negligible in comparison, a fingerprint of AMR.

The angular dependence of the resistance for magnetic field direction in the \(\beta\)-plane was studied using Hall bars oriented along the in-plane directions RuO\(_2\)[010] (\(J_C \perp c\)) and RuO\(_2\)[001] (\(J_C \parallel c\)), as shown in Fig.~\ref{fig_3}(a) and Fig.~\ref{fig_3}(b) for RuO\(_2\)[\SI{5}{\nano\meter}]/Py. Measurements were conducted at \SIlist{20;250;300}{\kelvin}. A strong angular dependence is evident that systematically decreases with increasing temperature, as illustrated in Fig.~\ref{fig_3}(c). An asymmetry in the angular dependence of resistivity is observed at $\beta = \pm \SI{90}{\degree}$ for both Hall bar orientations. Additionally, the resistance change in the \(\beta\)-plane exhibited minimal dependence on external magnetic field strength for RuO\(_2\)[\SI{5}{\nano\meter}]/Py, as shown in Fig.~\ref{fig_3}(d), highlighting weak magnetic field dependence.

\section{Discussion}
Our measurements reveal a clear dependence of the resistance on the magnetization orientation in both the $\alpha$- and $\gamma$-planes (Fig.~\ref{fig_2}(b)). Such behavior is consistent with the AMR in FMs, which arises from spin-orbit coupling and the resulting band splitting \cite{McGuirePotter1975}. In AMR, electron scattering is usually enhanced when the magnetization is parallel to the charge current compared to when it is perpendicular. This identical pattern is also observed in the bare Py thin film (Fig.~\ref{fig_2}(c)), aligning precisely with the theoretical predictions and establishing a baseline for the comparison of more complex heterostructures.

A significant difference is observed in the $\beta$-plane measurements of RuO$_2$/Py heterostructures compared to bare Py. ADMR for magnetic field direction in the $\beta$-plane  has been  extensively documented across various heterostructures, including FM/HM \cite{PhysRevLett.110.206601, PhysRevLett.117.116602, PhysRevB.87.224401, PhysRevLett.128.067201, kang2020negative, jia2020thickness, yang2018anomalous, PhysRevB.87.184421}, FM/Light metal (LM) \cite{PhysRevB.108.184422}, AFM/HM \cite{PhysRevB.97.014417}, and bare FM systems \cite{yang2018anomalous, Jia_2020,jia2020thickness}. The underlying mechanisms attributed to this ADMR include SMR \cite{PhysRevLett.110.206601, PhysRevB.87.224401, PhysRevB.87.184421}, Rashba-Edelstein magnetoresistance (REMR)  \cite{PhysRevLett.117.116602}, orbital Rashba-Edelstein magnetoresistance (OREMR) \cite{PhysRevLett.128.067201}, IGSC \cite{kang2020negative, PhysRevB.108.184422}, anomalous Hall magnetoresistance (AHMR) \cite{yang2018anomalous}, and the geometric size effect (GSE) \cite{kang2020negative, PhysRevB.108.184422}, depending on the system. In our measurements, the bare Py film exhibits negligible $\beta$-plane ADMR, whereas the RuO$_2$/Py bilayer shows a pronounced angular dependence in this plane (Fig.~\ref{fig_3}(a) and (b)). Typically, in FM/HM, FM/LM, or bare FM systems, a high-resistance ("positive") ADMR state is expected for out-of-plane magnetization when mechanisms such as SMR, REMR, OREMR, or AHMR are dominant. However, our results indicate a lower-resistance ("negative") ADMR state when the magnetization is oriented out-of-plane compared to the in-plane configuration. This discrepancy suggests that conventional SMR, REE, OREE, and AHMR cannot account for the observed ADMR. Negative ADMR for magnetic field direction in the $\beta$-plane is more commonly associated with AFM/HM systems due to SMR, FM/HM and FM/LM interfaces due to IGSC and GSE, and in bare FM films primarily due to GSE. To determine the most plausible explanation for our findings, we will further examine the roles of altermagnetic SMR, antiferromagnetic SMR, GSE, and IGSC in the RuO$_2$/Py system.

In addition to the overall negative ADMR trend, an asymmetry at $\pm \SI{90}{\degree}$ is discernible, which can originate from finite transverse resistivity contributions (e.g., AHE) into the longitudinal resistance due to the specific  finite contact misalignment in the Hall bars during patterning or manual connection to contact pads. The asymmetry corresponds to about a \SI{3}{\degree} tilt in longitudinal equipotential lines to the electric field, which is a plausible tilt due to misalignment.

\subsection{Excluding the conventional spin Hall magnetoresistance and certain alternative mechanisms}


Recent theoretical calculations suggest that RuO$_2$ can act as an altermagnet, providing a crystal-axis-dependent spin-polarized current when the transport direction is along [010] but not along [001] \cite{Smejkal2022_PhysRevX}. We expect the magnetic easy axis of the RuO$_2$ to be along the c-axis, which should not be significantly influenced by the comparably weak external magnetic field of \SI{0.95}{\tesla} used in our experiments \cite{Feng2022_NatElectron}. If altermagnetism was the main mechanism for the ADMR signal, we would expect a pronounced crystallographic dependence (i.e., “positive” ADMR along particular axes). However, the observed ADMR in our $\beta$-plane measurements is largely independent of the crystallographic orientation (compare Fig.~\ref{fig_3}(a) and (b)) and shows a consistent “negative” sign. This does not necessarily rule out altermagnetism but indicates that we do not observe the effects of spin-polarized transport induced by spin-polarized current from RuO$_2$. 


Negative ADMR in AFM/HM systems typically results from the interaction of a spin current driven by the SHE with the N\'eel order in the AFM. However, we do not observe any dependence on the c-axis (N\'eel order) by changing the Hall bar parallel or perpendicular to the c-axis \cite{manchon2017spin, PhysRevB.98.024422} or a significant correlation with magnetic field magnitude that would suggest domain realignment \cite{10.1063/1.4997588}, both of which would be anticipated in an AFM scenario. Nonetheless, a pronounced realignment of domains is not expected for RuO$_2$ at a magnetic field of \SI{5}{\tesla} \cite{tschirner2023saturation}. Furthermore, the temperature dependence of the magnetoresistance remains similar for two orthogonal crystallographic orientations, which further rules out the AFM-based explanation as we expect it to behave differently along N\'eel order and orthogonal to it as we change temperature \cite{PhysRevB.97.014417, 10.1063/1.4997588}.

The geometric size of a thin film can result in different scattering conditions based on whether the magnetization is in-plane or out-of-plane, sometimes leading to negative ADMR for magnetic field direction in the $\beta$-plane \cite{rijks1995semiclassical}. The relatively large magnitude of the negative ADMR observed in the RuO$_2$/Py sample, compared to a single Py film produced under similar conditions, exceeds the predictions of simple GSE models \cite{kang2020negative, PhysRevB.108.184422}. The texture and thickness of a thin film have been demonstrated to influence the GSE. For instance, a textured film has been shown to enhance ADMR, while a thicker film may diminish it in the $\beta$-plane \cite{PhysRevB.56.362}. In our study, both Py thin films were grown on substrates with very similar lattice constants, suggesting they possess comparable textures, which should result in similar GSE effects. Furthermore, the effective thickness of the conducting channel is increased for the bilayer compared to bare Py, showing higher negative ADMR. This outcome contradicts the anticipated decrease in negative ADMR with an increase in thickness due to GSE. Nevertheless, we observe significantly different GSE behaviors in the two samples, surpassing simple GSE explanations. Previous research has attempted to isolate GSE contributions by varying the thickness of the Py layer, demonstrating that GSE alone cannot account for the observed strong negative ADMR \cite{kang2020negative, PhysRevB.108.184422}.

\subsection{Interface-induced magnetoresistance}

Based on our prior discussion, it is evident that altermagnetic SMR, antiferromagnetic SMR, and GSE do not provide a satisfactory explanation for the observed ADMR. Therefore, it is necessary to consider an alternative mechanism that may account for this signal. The observations suggest IGSCs \cite{PhysRevLett.121.136805} at the RuO$_2$/Py interface could be the dominant source of the negative ADMR in the $\beta$-plane \cite{kang2020negative, PhysRevB.108.184422}. Interface-generated spin currents are spin currents generated from spin-orbit scattering within a mean free path of the interface. For example, an in-plane charge current \( J_C \) induces a perpendicularly-flowing spin current $Q_{zy}$ with flow direction $z$ (i.e. out-of-plane) and spin direction $y$ (parallel to the spin direction of the spin Hall current) due to spin-orbit scattering at the interface \cite{PhysRevB.94.104419, PhysRevLett.121.136805, PhysRevB.94.104420, PhysRevB.88.214417, fan2014quantifying}. The momentum-dependent and spin-dependent scattering processes that cause IGSCs are inherently captured in multilayer Boltzmann simulations. However, as shown in Refs. \cite{PhysRevB.94.104419, PhysRevB.94.104420}, IGSCs can also be captured by drift-diffusion models by inserting a spin current source term at the interface. Note that the bulk SHE is modeled using a similar source term in the bulk layers.

We now describe a mechanism in which the interface causes a longitudinal magnetoresistance: At the RuO$_2$/Py interface, the in-plane electric field generates a spin current $Q_{zy}$. This particular orientation of IGSC is created via the spin-orbit filtering process \cite{PhysRevLett.121.136805}. This spin current is converted to a longitudinal charge current via the iSHE. Even if the source term for the IGSC in the drift-diffusion equations is magnetization-independent, the boundary conditions at the interface---described by magnetoelectronic circuit theory---are magnetization-dependent. As a result, the total spin current that forms in the bulk nonmagnetic layer depends on whether \( M \parallel y \) or \( M \perp y \). Thus, the contribution to the longitudinal charge current from the iSHE will be different for \( M \parallel y \) or \( M \perp y \), resulting in a contribution to the ADMR. Our drift-diffusion calculations show that the total spin current in the nonmagnetic layer (and thus the total backflow of longitudinal charge current) is greater when \( M \parallel y \) versus \( M \parallel z \) for a large parameter space, indicating that ``negative" ADMR can result from the combination of IGSCs and the iSHE.

Before continuing, we note previous work by Kang et al. \cite{kang2020negative} which studies the longitudinal magnetoresistance versus magnetic field direction in the $\beta$-plane using a spin drift-diffusion model \cite{PhysRevB.87.144411} that includes IGSCs \cite{PhysRevB.94.104419, PhysRevLett.121.136805, PhysRevB.94.104420, PhysRevB.88.214417} and utilizes symmetry arguments based on Onsager reciprocity. In their model, the contribution from IGSCs and the iSHE effect is canceled by a reciprocal contribution. In the reciprocal contribution, the SHE first generates an out-of-plane flowing spin current and the interface converts it into an in-plane charge current via spin-orbit scattering. One could refer to this as interfacial spin-to-charge conversion or an \emph{inverse} IGSC. We caution that the Onsager reciprocity of IGSCs has not been confirmed and is unlikely to hold due to the breaking of time-reversal symmetry at the interface. Thus, we do not expect these contributions to cancel in realistic systems. The authors further introduce an entirely interfacial magnetoresistance due to both IGSCs and their inverse counterpart. Such contributions are allowed by symmetry but should be studied entirely in the context of the Boltzmann formalism, since the conversion processes take place entirely within a mean free path from the interface. We leave such calculations for future work, as they are not required to provide a qualitative explanation of a negative ADMR.

Thus, in this work, we focus on the interplay between IGSCs and the iSHE because it is the simplest way to show how negative ADMR can result from interfacial scattering. Figure~\ref{fig_3}(e) and (f) illustrate how the combination of IGSCs and the iSHE lead to a greater longitudinal resistance for \( M \parallel y \) compared to \( M \parallel z \). Here, \( J_C \) represents the applied charge current, while \( J_C^{\text{ind}} \) corresponds to the induced charge current generated via the spin-to-charge conversion processes. The net spin current \( Q_{zy}^{\text{net}} \) at the RuO$_2$/FM interface follows the relation \( Q_{zy}^{\text{net}}(M = z) < Q_{zy}^{\text{net}}(M = y) \), since the absorption of spin current by the FM layer \( Q_{zy}^{\text{abs}} \) is larger when \( M \) is aligned along the \( z \)-axis compared to the \( y \)-axis. Unlike contributions from the bulk SHE and AHE, for which spin currents are generated away from the interface and are partially reflected at the RuO$_2$/Py boundary, the IGSCs originate directly at the interface and are not subject to such reflections \cite{kang2020negative, PhysRevLett.121.136805}. This mechanism can account for the observed lower resistance when the magnetization is out-of-plane, resulting in the negative ADMR state. The temperature dependence of this effect further supports this interpretation, as elevated temperatures reduce coherent spin-dependent scattering and decrease the mean-free path, thereby diminishing the contribution from IGSC and weakening the ADMR (Fig.~\ref{fig_3}(d)). Additionally, the independence of ADMR from the applied magnetic field strength suggests that the IGSCs are primarily governed by the magnetization direction rather than the magnitude of the external magnetic field, which explains the lack of significant magnetic field-strength influence in our data (Fig.~\ref{fig_3}(c)).

The spin drift-diffusion equations are given by
\begin{align}
\boldsymbol{\mu}(z) &=
  \begin{cases} 
   -\frac{\tau^\text{NM}_\text{SF}}{e} \frac{\partial \boldsymbol{Q}_z}{\partial z} & z < 0 \\
   -\frac{\tau^\text{FM}_\text{SF}}{e} \frac{\partial \boldsymbol{Q}_z}{\partial z} & z > 0
  \end{cases} \\
\boldsymbol{Q}_z(z) &=
  \begin{cases} 
   -\frac{\sigma_\text{NM}}{2e} \frac{\partial \boldsymbol{\mu}}{\partial z} + Q_\text{SHE}\boldsymbol{\hat{y}} & z < 0 \\
   -\frac{\sigma_\text{FM}}{2e} \frac{\partial \boldsymbol{\mu}}{\partial z} & z > 0
  \end{cases}
\end{align}
with $\boldsymbol{\mu}$ and $\boldsymbol{Q}_z$ being the spin accumulation and z-flowing spin current respectively. Note that the components of the spin current are given by $[\boldsymbol{Q}_z]_i = Q_{zi}$ for $i \in [x,y,z]$. The parameters $\tau^{\text{NM/FM}}_\text{SF}$ and $\sigma_{\text{NM/FM}}$ are the spin flip lifetimes and the bulk conductivities respectively, and can be related to the spin diffusion lengths $l_\text{NM/FM}$. The interface is located at $z = 0$ with the negative $z$ region belonging to the NM and the positive $z$ region belonging to the FM.

We assume that the z-flowing charge current vanishes everywhere and all spin currents vanish at the outer layer boundaries ($z = -t_\text{NM}$ and $z = t_\text{FM}$). At the NM|FM interface, the boundary conditions are given by magnetoelectronic circuit theory with an additional source ($Q_{IGSC}$) capturing the IGSC. Note that we assume a positive value of $Q_{IGSC}$, which as we shall see, leads to a mostly negative ADMR over the swept parameter space. A negative value of $Q_{IGSC}$ would reverse this trend and result in a primarily positive ADMR. In each case, however, the sign of the ADMR is not fixed and flips for certain layer thicknesses. Thus, the proposed interfacial mechanism can result in negative ADMR for both positive and negative values of $Q_{IGSC}$. Realistic calculations of the sign and magnitude of $Q_{IGSC}$, which require careful first-principles transport calculations, are outside the scope of this work.

When $M \parallel z$, the interface boundary conditions are:
\begin{widetext}
\begin{align}
 \begin{pmatrix}
  Q_{zx}(0^-) \\
  Q_{zy}(0^-) \\
  Q_{zz}(0) \\
  Q_{zc}(0)
 \end{pmatrix} = 
  \frac{1}{e}
 \begin{pmatrix}
  \Re[G_{\uparrow\downarrow}] & -\Im[G_{\uparrow\downarrow}] & 0 & 0 \\
  \Im[G_{\uparrow\downarrow}] & \Re[G_{\uparrow\downarrow}] & 0 & 0 \\
  0  & 0  & G_+ & G_-  \\
  0 & 0 & G_- & G_+ 
 \end{pmatrix}
 \begin{pmatrix}
  \mu_x(0^-) \\
  \mu_y(0^-) \\
  \mu_z(0^+) - \mu_z(0^-) \\
  \mu_c(0^+) - \mu_c(0^-)
 \end{pmatrix}
 +
  \begin{pmatrix}
  0 \\
  Q_{IGSC} \\
  0 \\
  0
 \end{pmatrix}
 \label{BCmz}
\end{align}
\end{widetext}
Likewise, when $M \parallel y$, the interface boundary conditions are:
\begin{widetext}
\begin{align}
 \begin{pmatrix}
  Q_{zx}(0^-) \\
  Q_{zy}(0) \\
  Q_{zz}(0^-) \\
  Q_{zc}(0)
 \end{pmatrix} =
  \frac{1}{e}
 \begin{pmatrix}
  \Re[G_{\uparrow\downarrow}] & 0 & -\Im[G_{\uparrow\downarrow}] & 0 \\
  0 & G_+ & 0 & G_-   \\
  \Im[G_{\uparrow\downarrow}] & 0  & \Re[G_{\uparrow\downarrow}] & 0 \\
  0 & G_- & 0 & G_+ 
 \end{pmatrix}
 \begin{pmatrix}
  \mu_x(0^-) \\
  \mu_y(0^+) - \mu_y(0^-) \\
  \mu_z(0^-) \\
  \mu_c(0^+) - \mu_c(0^-)
 \end{pmatrix}
 +
  \begin{pmatrix}
  0 \\
  Q_{IGSC} \\
  0 \\
  0
 \end{pmatrix}
 \label{BCmz}
\end{align}
\end{widetext}
These boundary conditions capture several well-known features of interfacial spin transport, such as dephasing of spins transverse to the magnetization and conservation of spins longitudinal to the magnetization. Note that $\Re[G_{\uparrow\downarrow}]$ and $\Im[G_{\uparrow\downarrow}]$ are the real and imaginary parts of the spin mixing conductance respectively and $G_\pm = G_\uparrow \pm G_\downarrow$ are the interfacial conductance parameters governing charge and longitudinal spin transport. To good approximation, the boundary conditions are split into two independent regimes governing 1) the transverse spin components and the 2) longitudinal spins and charge components \cite{PhysRevB.94.104419,PhysRevB.94.104420}. Since both the SHE and the spin-orbit filtering current have spin direction along $y$, these spin currents will dephase in the FM when $m \parallel z$ and transmit through the FM when $m \parallel y$. According to the drift-diffusion solution, spin currents generated at the interface behave differently than spin currents generated in the bulk.

Note that the consequences of interfacial spin-orbit coupling on magnetoelectronic circuit theory have been extensively studied \cite{PhysRevB.94.104419,PhysRevB.94.104420, PhysRevB.101.224405, PhysRevLett.117.207204, PhysRevLett.116.196602}. With interfacial spin-orbit coupling, all elements of the matrix in Eq.~\ref{BCmz} are nonvanishing; however, to good approximation the off-diagonal elements can be treated as zero. The main contribution of interfacial spin-orbit coupling is (longitudinal) spin memory loss and IGSCs. For simplicity we only include the latter here; we do not expect the presence of longitudinal spin memory loss to qualitatively change our results. The important result is that the boundary conditions are magnetization-dependent even if the source of the spin current is not, which causes the total spin current in the NM to be magnetization dependent, ultimately causing a nonzero magnetoresistance.

\begin{figure*}
    \centering
    \includegraphics[width=\textwidth]{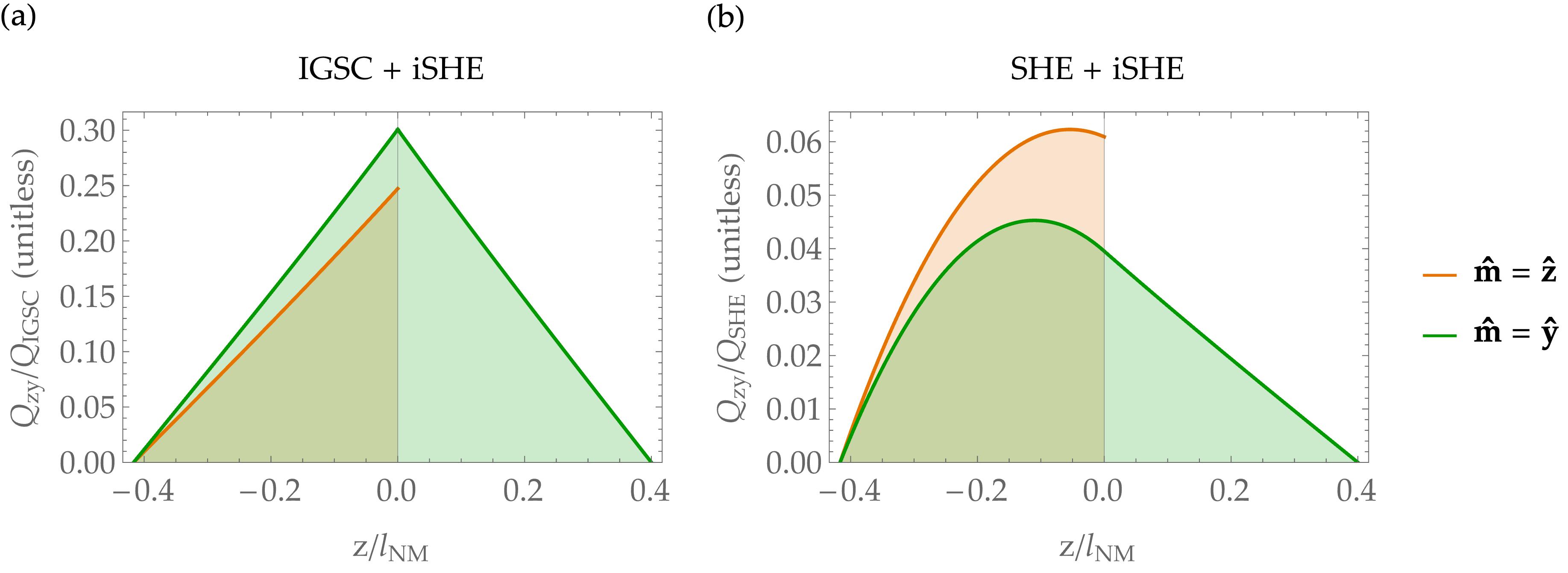}
    \caption{Plots of $Q_{zy}$ versus $z$ for the (a) interfacial and (b) spin Hall contribution to the longitudinal magnetoresistance. Each case considers a different source of $Q_{zy}$ (i.e. interfacial versus bulk), but in both cases, the iSHE in the NM layer is assumed to convert $Q_{zy}$ to an in-plane charge current $j_x$, which opposes the original charge current from the applied electric field. Panel (a) shows that $Q_{zy}$ is greatest for $m \parallel y$, indicating a negative ADMR from the interfacial contribution. Panel (b) shows that $Q_{zy}$ is greatest for $m \parallel z$, indicating a positive ADMR from the bulk spin Hall contribution.} 
    \label{fig_zplots}
\end{figure*}

\begin{figure*}
    \centering
    \includegraphics[width=\textwidth]{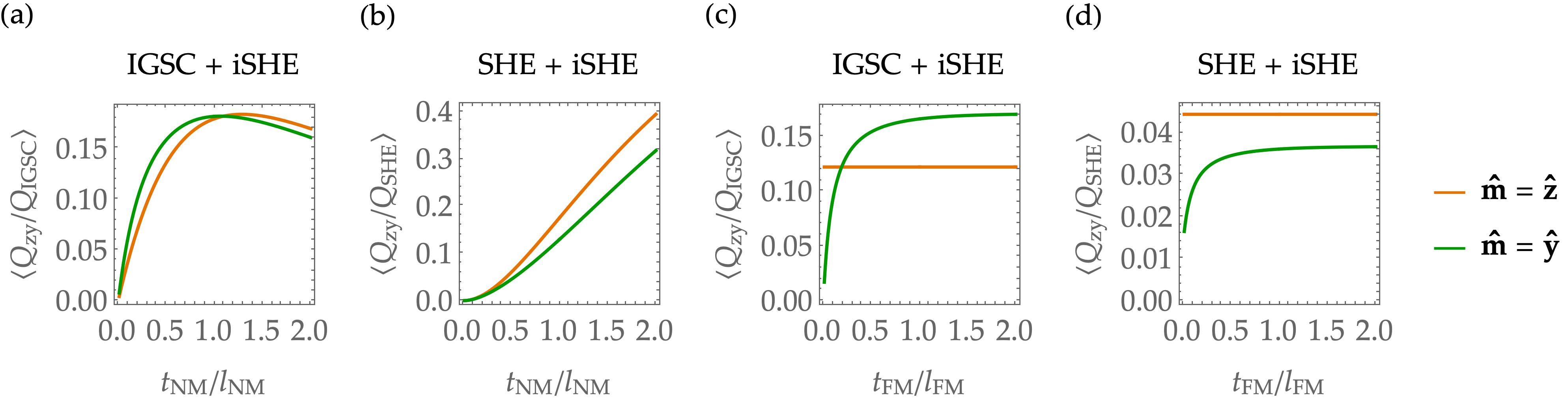}
    \caption{Thickness dependencies of the average, dimensionless spin current in the NM layer, obtained by integrating $Q_{zy}$ from $z = -t_\text{NM}$ to $z = 0$ and dividing by $t_\text{NM}$ and the source spin current strength. Panels (a) and (c) show that the interfacial contribution to the MR ratio can switch from negative to positive in certain regimes.}
    \label{fig_thickness}
\end{figure*}

Fig.~\ref{fig_zplots} plots $Q_{zy}$ versus out-of-plane position $z$ when the only spin current source is either the (a) IGSC or (b) the spin Hall current. Panel (a) shows that the IGSC diffuses away from the interface into the bulk layers, while panel (b) shows the spin Hall current generated in the bulk of the NM layer diffuses through the interface into the FM layer. In the NM layer, the area under the curve is proportional to the total spin current in that layer, and thus proportional to the backflow of charge current from the iSHE. Clearly, for (a), the spin current in the NM layer is greatest when \( M \parallel y \), resulting in the greatest charge backflow and a negative ADMR. In contrast, in (b) the spin current in the NM layer is smallest when \( M \parallel y \), resulting in the least charge backflow and a positive ADMR. Figure~\ref{fig_thickness}(a)-(b) shows the average spin current $Q_{zy}$ in the NM layer as a function of $t_\text{NM}$ for $t_\text{FM} = \SI{4}{\nano\meter}$. For smaller thicknesses, the interface-induced mechanism results in $Q_{zy}(m \parallel y) > Q_{zy}(m \parallel z)$, indicating a negative ADMR, while the spin Hall-induced mechanism results in the opposite trend (positive ADMR). Note, however, that the interfacial contribution becomes positive for larger thicknesses. Figure~\ref{fig_thickness}(c)-(d) shows the average spin current in the NM as a function of $t_\text{FM}$ for $t_\text{NM} = \SI{5}{\nano\meter}$. Here we see the same trends as before, except $Q_{zy}(m \parallel z)$ does not depend on $t_\text{FM}$. This is because, when $m \parallel z$, the spin current $Q_{zy}$ dephases within a few atomic layers from the interface and is thus insensitive to $t_\text{FM}$.

\begin{figure*}
    \centering
    \includegraphics[width=\textwidth]{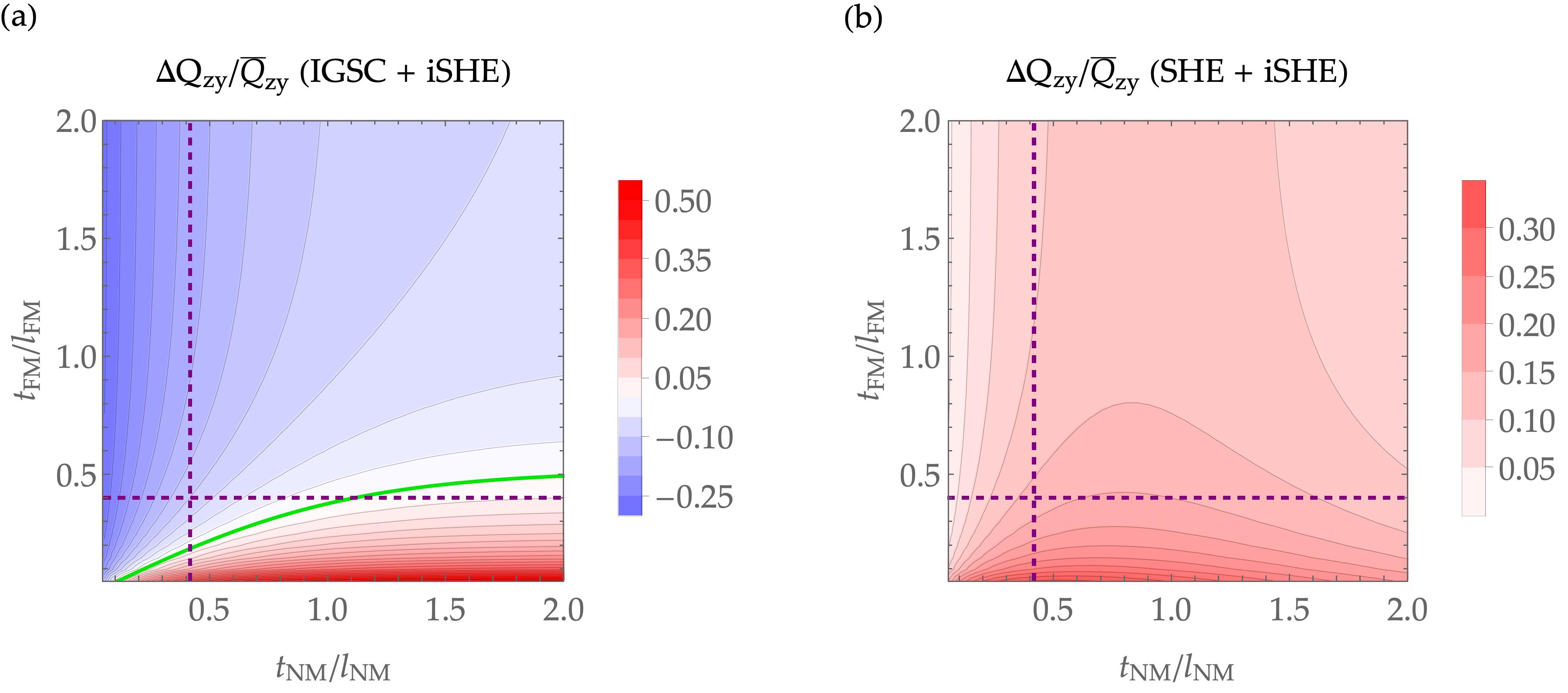}
    \caption{Contour plot of $\Delta Q_{zy}/\Bar{Q}_{zy}$ from both (a) interfacial and (b) bulk spin Hall contributions as a function of $t_\text{NM}$ and $t_\text{FM}$. The bulk contributions are always positive while the interfacial contributions are mostly negative. The threshold between negative and positive values, given by the green line, can be calculated analytically using Eq.~\ref{crossover}. Note that negative values of $\Delta Q_{zy}/\Bar{Q}_{zy}$ correspond to $Q_{zy}(m \parallel z) < Q_{zy}(m \parallel y)$, or ``negative" ADMR. The horizontal purple dashed lines indicate the plotted region in Fig.~\ref{fig_thickness}(a)-(b) and the vertical purple dashed lines indicate the plotted region in Fig.~\ref{fig_thickness}(c)-(d). The green solid line marks the transition from positive to negative ADMR as is determined by Eq.~\ref{crossover}.
}
    \label{fig_contour}
\end{figure*}

Figure~\ref{fig_contour} shows contour plots of the scaled difference between $Q_{zy}(m \parallel z)$ and $Q_{zy}(m \parallel y)$, defined as:
\begin{align}
\frac{\Delta Q_{zy}}{\Bar{Q}_{zy}} &= \frac{Q_{zy}(m \parallel z) - Q_{zy}(m \parallel y)}{Q_{zy}(m \parallel z) + Q_{zy}(m \parallel y)},
\end{align}
plotted versus the scaled thicknesses $t_\text{NM}/l_\text{NM}$ and $t_\text{FM}/l_\text{FM}$. Since ``positive" ADMR means that $Q_{zy}(m \parallel z) > Q_{zy}(m \parallel y)$ and ``negative" ADMR means that $Q_{zy}(m \parallel z) < Q_{zy}(m \parallel y)$, the sign of $\Delta Q_{zy}/\Bar{Q}_{zy}$ is also the sign of the ADMR. Panel~\ref{fig_contour}(a) shows that the ADMR caused by IGSC and the iSHE is negative for a large portion of parameter space, while Panel~\ref{fig_contour}(b) shows that the ADMR caused by the SHE and iSHE (i.e. the conventional SMR) is always positive.

The conditions for a negative, interfacial ADMR in the limit of vanishing imaginary part of the spin mixing conductance (i.e. $\Im[G_{\uparrow\downarrow}] = 0$) is given by the following inequality,
\begin{align}
\frac{\Re[G_{\uparrow\downarrow}]}{G_\parallel} > 2 \Bigg{(} \frac{l_\text{FM}}{l_\text{NM}} \frac{\sigma_\text{NM}}{\sigma_\text{FM}} \frac{\coth(t_\text{FM}/l_\text{FM})}{\coth(t_\text{NM}/l_\text{NM})} + 1 \Bigg{)}
\label{crossover}
\end{align}
with
\begin{align}
G_\parallel \equiv \Big{(} \frac{1}{G_\uparrow} + \frac{1}{G_\downarrow} \Big{)}^{-1}.
\end{align}
When the L.H.S. and R.H.S. of Eq.~\ref{crossover} are equal, one obtains a transcendental equation for the crossover from negative to positive interfacial ADMR, which is the green curve in panel~\ref{fig_contour}(a). Unlike the interfacial contribution, panel~\ref{fig_contour}(b) shows that the spin Hall contribution is always positive, regardless of layer thickness.

\subsection{Other interpretations}

This understanding is built upon existing theoretical and experimental studies of FM/HM and FM/LM interfaces, emphasizing the importance of the interface’s electronic structure \cite{kang2020negative, PhysRevB.108.184422}. Whether the orbital angular moment in RuO$_2$ further enhances IGSC at the interface remains an open question for future investigation \cite{yahagi2024neel}. While we have carefully ruled out dominant altermagnetism, AFM order in RuO$_2$, and GSE as the primary mechanisms, the subtle interplay between these effects and IGSC may still require further investigation. For instance, a more detailed characterization of the interface structure using transmission electron microscopy could provide insights into the extent of hybridization at the RuO$_2$/Py interface. A systematic study of layer thickness variations in RuO$_2$ could reveal further details on the spin transmission and reflection mechanisms at the interface. Furthermore, investigating the role of orbital contributions, specifically how orbital angular momentum in RuO$_2$ influences IGSC, may offer additional control parameters for tuning spin current generation via orbital effects for magnetotransport behavior.

Recent experimental studies on RuO$_2$/FM systems with RuO$_2$(101) orientation have provided new insights into the SMR in altermagnetic materials \cite{pan2024unveiling, chen2024altermagnetic}. These studies have observed spin-polarized currents in both orthogonal Hall bars within the thin film plane, accompanied by a tilting of the spin current relative to the film plane. This finding is particularly relevant as it suggests additional complexity in the spin transport mechanisms within altermagnetic heterostructures, complementing the theoretical expectations. Our study further explores how crystal orientation influences spin current polarization in RuO$_2$(100)/FM heterostructures. Our results highlight the dominant role of IGSC and provide a foundation for further studies to disentangle interfacial and bulk contributions.

\section{Conclusion}
In summary, we have systematically investigated RuO$_2$/Py heterostructures, demonstrating a robust negative ADMR for magnetic field direction in the $\beta$- plane. Our results provide compelling evidence that IGSC at the RuO$_2$/Py interface governs the observed magnetotransport effects, overshadowing any potential effects of spin-polarized transport induced by spin-polarized current from RuO$_2$. We also established the role of IGSC in combination with iSHE, which leads to observed magnetoresistance with the drift-diffusion calculations. This finding underscores the critical role of interfacial phenomena in oxide/FM systems and supports the importance of further refined experiments to isolate subtle altermagnetic signals. While our data do not rule out the existence of magnetism in RuO$_2$, they emphasize the necessity of additional investigations, possibly with modified device geometries or advanced measurement techniques, to identify possible contributions. 
Ultimately, our study neither fully confirms nor rules out RuO$_2$ altermagnetism but highlights the often neglected powerful influence of IGSC in governing magnetoresistance behaviors. These findings provide new insights into interfacial spin transport mechanisms, which could be leveraged for engineering IGSC-driven spintronic memory and sensing devices with enhanced efficiency.

\begin{acknowledgments}

The authors acknowledge Sachin Krishnia for valuable scientific discussions that enriched the insights of this work. All authors from Mainz also gratefully acknowledge funding support from the Deutsche Forschungsgemeinschaft (DFG) under the framework of the Collaborative Research Center TRR 173–268565370 Spin+X (Project B02 and A01) and TRR 288-268565370 Elasto-Q-Mat (Project A12). V.A. and E.M. were supported by the National Science Foundation
under grant ECCS-2236159. These contributions have been instrumental in enabling this research. V.A. and E.M. performed the theoretical calculations and wrote the sections pertaining to those calculations. The authors declare no conflicts of interest.

\end{acknowledgments}

\section*{Data Availability}
The data supporting the findings of this study are openly available in Zenodo \cite{akashdeep_2025_17315784}.

%

\end{document}